\documentclass[aps,prl,twocolumn,showpacs,groupedaddress]{revtex4}

\usepackage{amsmath,amssymb,graphicx,color}
\usepackage{bm}

\begin{document}

\title{Unconventional photon blockade in coupled optomechanical systems}

\author{Vincenzo Savona}
\email{vincenzo.savona@epfl.ch}
\affiliation{Institute of Theoretical Physics, Ecole Polytechnique F\'{e}d\'{e}rale de Lausanne (EPFL), CH-1015 Lausanne, Switzerland}

\date{\today}

\begin{abstract}
We show that in laser-driven coupled optomechanical systems, photon antibunching can occur under weak optomechanical coupling, contrarily to common expectation. This unconventional photon blockade originates from destructive quantum interference between different excitation pathways bringing from the ground to two-photon states. Using a quantum open-system approach, we study the antibunching as a function of driving field amplitude, temperature, and pure dephasing rate, and derive optimal values of the system parameters for its occurrence. These values are remarkably close to those characterizing optomechanical systems in some current experimental studies.
\end{abstract}

\maketitle

Optomechanics is the study of optical cavities in which the resonant mode is coupled to a mechanical oscillator by means of radiation pressure. The field has experienced a tremendous progress in recent years \cite{Kippenberg2008,Marquardt2009,Meystre2012}, marked by several achievements, among which the successful optical cooling close to the ground state of the mechanical motion \cite{Chan2011,Teufel2011,Verhagen2012} and the coherent coupling between the two degrees of freedom \cite{Verhagen2012}. In addition to the promise of operating optomechanical systems (OMSs) as extremely sensitive detectors of mass, force or position, perhaps the most interesting perspective is the realization of strong single-photon nonlinearities. These would bring to several possible schemes for producing quantum correlated states of light and macroscopic mechanical degrees of freedom, with possible applications in quantum information processing \cite{Rabl2010,Stannigel2012}. 

An OMS is predicted to display single-photon nonlinearities \cite{Kronwald2013,Ludwig2012,Nunnenkamp2011,Rabl2011,Stannigel2012}, provided the optomechanical coupling is strong enough to overcome system losses -- in analogy to what happens in cavity quantum electrodynamics (CQED) \cite{Carmichael2008}. In particular, in both single \cite{Kronwald2013,Rabl2011} and coupled \cite{Ludwig2012,Stannigel2012} OMSs, the emission of photons with sub-poissonian statistics is expected. This kind of {\em photon blockade} is what occurs for a resonantly driven two-level system \cite{Carmichael1999}: after emission of one photon, the system needs to be excited again and a second photon emission is most likely to be delayed by one lifetime. Any system with a strongly anharmonic set of energy levels, when driven resonantly to one of its possible optical transitions, will essentially behave in the same fashion.

A mechanism of completely different nature however -- that has been recently named \cite{Carusotto2013} {\em unconventional photon blockade} (UPB) as opposed to the conventional mechanism (CPB) described above -- can give rise to sub-poissonian photon emission in CQED even in presence of much weaker nonlinearities. In this mechanism, the presence of two photons in the system is prevented by destructive quantum interference -- enforced by the small nonlinearity -- between different excitation pathways that can lead from the ground to the two-photon state. UPB has been initially predicted by Carmichael \cite{Carmichael1985} for the Jaynes-Cummings model and successfully demonstrated in atomic CQED experiments \cite{Foster2000,Rempe1991}. More recently, UPB was predicted \cite{Bamba2011,Liew2010} in coupled CQED systems in presence of very small nonlinearities -- both of the Kerr and Jaynes-Cummings types -- such that the phenomenon might even be induced by the background material nonlinearity of modern semiconductor nanocavities \cite{Ferretti2013}.

In this letter, we investigate the occurrence of UPB in coupled OMSs, and characterize it as a function of losses, driving field strength and dephasing mechanisms. We show that UPB is expected to occur well in the weak optomechanical coupling limit, contrarily to CPB. By reviewing the state-of-the-art parameters of current OMSs, we show that some of them are close to the UPB requirements and set up a perspective for the experimental observation of this phenomenon.

Consider two linearly coupled optical modes, described by bosonic creation operators $\hat{a}_1^\dagger$ and $\hat{a}_2^\dagger$. Mode two is characterized by optomechanical coupling to a phonon mode described by a bosonic creation operator $\hat{b}_2^\dagger$. Mode one is driven by a monochromatic laser field at frequency $\omega_p$. In the frame rotating with the driving frequency, the Hamiltonian is (assuming $\hbar=1$)
\begin{eqnarray}
\hat{H}&=&\Delta_1\hat{a}^\dagger_1\hat{a}_1+\Delta_2\hat{a}^\dagger_2\hat{a}_2+\omega_m\hat{b}^\dagger_2\hat{b}_2\notag\\
&-&J(\hat{a}^\dagger_1\hat{a}_2+\hat{a}^\dagger_2\hat{a}_1)
+g\hat{a}^\dagger_2\hat{a}_2(\hat{b}^\dagger_2+\hat{b}_2)+\epsilon(\hat{a}^\dagger_1+\hat{a}_1)\,,\label{eq:OmHam}
\end{eqnarray}
where $\Delta_1=\omega_1-\omega_p$ and $\Delta_2=\omega_2-\omega_p$ are the detunings of the two mode frequencies, $J$ is the rate of the linear coupling, $\omega_m$ is the mechanical frequency, $g$ is the single-photon optomechanical coupling rate,  and $\epsilon$ denotes the (real) amplitude of the driving field. We further assume dissipation rates $\kappa_{1,2}$ and $\gamma$ for the optical and mechanical modes respectively. 

The dynamics of this driven-dissipative system is governed by the Von-Neumann equation for the density matrix, $\hat{\rho}(t)$:
\begin{eqnarray}
\frac{d \hat{\rho}}{dt}&=&-i\left[\hat{H},\hat{\rho}\right]+\sum_{j=1,2}\frac{\kappa_j}{2}\mathcal{D}[\hat{a}_j]\hat{\rho}\notag\\
&+&\frac{\gamma}{2}\left[(N_{th}+1)\mathcal{D}[\hat{b}_2]\hat{\rho}+N_{th}\mathcal{D}[\hat{b}^\dagger_2]\hat{\rho}\right]\,,\label{eq:master}
\end{eqnarray}
where $\mathcal{D}[\hat{O}]\hat{\rho}=2\hat{O}\hat{\rho}\hat{O}^\dagger-\hat{O}^\dagger\hat{O}\hat{\rho}-\hat{\rho}\hat{O}^\dagger\hat{O}$ are Lindblad terms describing the markovian interaction with the environment and $N_{th}=(e^{\omega_m/K_BT}-1)^{-1}$ is the average occupation number of the mechanical mode for a given temperature $T$. 

In the absence of driving terms, the photon and phonon degrees of freedom in an OMS can be decoupled by means of a polaron transformation \cite{Rabl2011}. As a result, the optomechanical coupling term in (\ref{eq:OmHam}) is replaced by $\hat{H}^\prime=-\Delta_g(\hat{a}_2^\dagger\hat{a}_2+\hat{a}_2^\dagger\hat{a}_2^\dagger\hat{a}_2\hat{a}_2)$, with $\Delta_g=g^2/\omega_m$, representing a polaron energy shift and an effective Kerr-type nonlinearity on mode two. This sets a close analogy between the present system and the weakly nonlinear coupled optical modes studied in Refs \cite{Liew2010,Bamba2011}, that were shown to display UPB under appropriate conditions on the system parameters. We thus set out to demonstrate that Hamiltonian (\ref{eq:OmHam}) can also give rise to UPB and investigate its features. 

We assume $\kappa_1=\kappa_2=\kappa$, and use this quantity to rescale all other energy parameters. The UPB arises in presence of a weak nonlinearity in the second optical mode, which amounts here to choosing $\Delta_g/\kappa\ll1$. It further requires $\omega_m/\kappa\gg1$ -- namely the well-resolved sideband limit -- as explained below. These two conditions naturally allow for a weak optomechanical coupling $g/\kappa\lesssim1$. We consider three fixed values of the mechanical frequency: $\omega_m=11\kappa$, $24\kappa$, and $50\kappa$, characterizing the state of the art in different kinds of OMSs \cite{Chan2011,Chan2012,Verhagen2012}. UPB depends critically on the value of $\Delta_2$ (while being much less sensitive to $\Delta_1$ \cite{Liew2010,Ferretti2013}). Hence, throughout this work, we set the value of the polaron-shifted detuning $\Delta_2-\Delta_g$ to the optimal UPB condition derived in Ref. \onlinecite{Bamba2011} in the $\epsilon\to0$ limit, $\Delta_{opt}=-(1/2)\sqrt{\sqrt{9J^2+8\kappa^2J^2}-\kappa^2-3J^2}$. We additionally set $\Delta_1=\Delta_{opt}$ and $\gamma=0.01\kappa$, and study the result as a function of the remaining parameters.

\begin{figure}
\includegraphics[width=0.5 \textwidth]{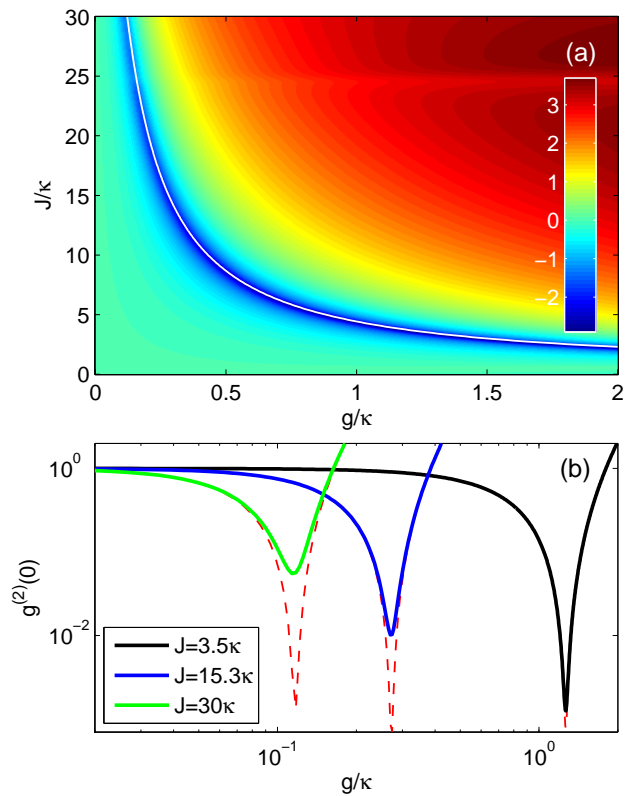}
\caption{\label{fig1} (a) $\log_{10}(g^{(2)}(0))$ computed for $\omega_m=50\kappa$ and $\epsilon=0.1\kappa$, as a function of $g$ and $J$. White line: optimal UPB condition given by Eq. (5) of Ref. \onlinecite{Bamba2011} (b) Three horizontal cuts of the color-plot in (a) for the values of $J$ indicated in the legend. Dashed lines: same quantity computed by restricting to the zero-phonon subspace.}
\end{figure}

For the numerical analysis, we assume steady state, $\dot{\hat{\rho}}=0$, and solve the linear system arising from (\ref{eq:master}) under the condition $\mbox{Tr}(\hat{\rho})=1$, by truncating the Hilbert space of the system to the basis $\{|n_1,n_2,n_m\rangle\}$ with $n_1+n_2\leq N_{ph}$ and $n_m\leq N_m$. The numerical solution is obtained by means of a highly-optimized numerical code (see supplemental material). We have numerically checked that our results are scarcely affected by the assumption of an additional mechanical mode $\hat{b}_1$ coupled to the first optical mode. This is expected, as the photon occupation is much larger in mode two \cite{Liew2010,Bamba2011}. For all results, convergence vs $N_{ph}$ and $N_m$ was accurately verified.

In the present scheme, we expect the occurrence of photon antibunching in the first mode. We therefore compute the quantity $g^{(2)}(0) = \langle \hat{a}_1^\dagger\hat{a}_1^\dagger\hat{a}_1\hat{a}_1\rangle/\langle\hat{a}_1^\dagger\hat{a}_1\rangle^2$, where the quantum expectation value is evaluated as $\langle\hat{O}\rangle=\mbox{Tr}(\hat{\rho}\hat{O})$. Fig. \ref{fig1}(a) is a color plot of $\log_{10}(g^{(2)}(0))$ as a function of $g$ and $J$, for $\omega_m=50\kappa$ and a weak probe amplitude $\epsilon=0.1\kappa$. For most values of the coupling rate $J$, antibunching occurs in a well defined range of values of the optomechanical coupling (blue in the plot), with $g^{(2)}(0)$ taking values as small as $10^{-3}$. The white curve in the plot represents the optimal UPB condition on $\Delta_g$ derived by Bamba et al. for the case with Kerr nonlinearity \cite{Bamba2011} (see Eq. (5)). The agreement with our result confirms that the antibunching is indeed produced by the same UPB mechanism. Fig. \ref{fig1}(b) shows $g^{(2)}(0)$ as a function of $g$ for three selected values of $J$. We observe that a significant antibunching can be achieved for values of the optomechanical coupling as small as $g=0.1\kappa$. 

\begin{figure}
\includegraphics[width=0.45 \textwidth]{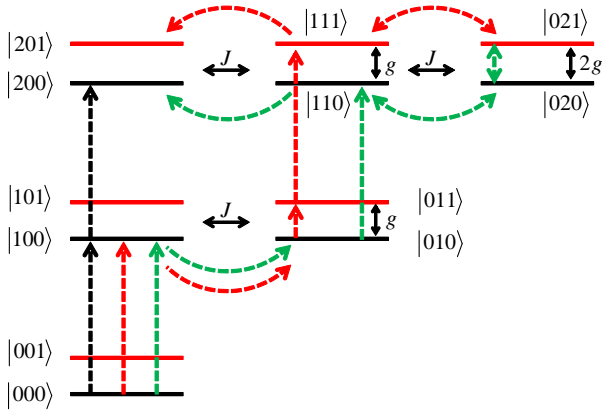}
\caption{\label{fig2} Sketch of the energy levels corresponding to Fock states up to $n_{ph}=2$ and $n_m=1$. Black (red) horizontal lines denote zero (one) phonon states. Dashed arrows indicate possible excitation pathways to states $|200\rangle$ (black and green) and $|201\rangle$ (red).}
\end{figure}

The present result differs however substantially from the one with Kerr nonlinearity in that the minimal value of $g^{(2)}(0)$ for a given $J$ is subject to a lower bound. In the Kerr case, for the optimal choice of the parameters, $g^{(2)}(0)$ decreases with the intensity of the driving field and can become arbitrarily small \cite{Carmichael1985,Liew2010,Bamba2011}. To understand this behaviour, we study the limit of vanishing $\epsilon$ by restricting the space of relevant Fock states $|n_1,n_2,n_m\rangle$ to $N_{ph}=2$ and $N_m=1$ (implying a total of 12 states). The corresponding bare energy levels are sketched in Fig. \ref{fig2} together with the relevant coupling terms. In order to achieve UPB, the amplitude of both the $|200\rangle$ and $|201\rangle$ states must be suppressed. In the case with Kerr nonlinearity, the $|200\rangle$-amplitude is suppressed via destructive interference between the linear excitation pathway (sketched as black dashed arrows) and all those involving the nonlinearity arising from an even number of phonon creation/destruction processes (e.g. the green pathway in the sketch). This condition is achieved for the optimal conditions derived in Ref. \onlinecite{Bamba2011}. In the present case however, there are also excitation pathways leading to the $|201\rangle$ state, involving an odd number of phonon creation/destruction processes. The same optimal conditions cannot suppress both amplitudes simultaneously, which ultimately explains the lower bound on $g^{(2)}(0)$ observed in the numerical results. The existence of this lower bound can be proved analytically by studying the Schr\"odinger equation on the 12-state basis described above (see supplemental material). To further support this picture, using the full model (\ref{eq:master}), we evaluate the second order photon correlation under the condition of zero phonon occupation. This can be defined using the operator $\hat{P}_0=\hat{I}_1\otimes\hat{I}_2\otimes|0_m\rangle\langle0_m|$, that projects on the zero-phonon subspace: $g^{(2)}_0(0)\equiv\langle \hat{a}_1^\dagger\hat{a}_1^\dagger\hat{a}_1\hat{a}_1\hat{P}_0\rangle/\langle\hat{a}_1^\dagger\hat{a}_1\hat{P}_0\rangle^2$. For each of the three sets of parameters considered in Fig. \ref{fig1}(b), this quantity is plotted as dashed red lines. The lower bound on $g^{(2)}(0)$ is clearly suppressed and the minimal value of $g^{(2)}_0(0)$ is now set only by the finite value of the driving field $\epsilon$ used in the numerical calculation.

\begin{figure}
\includegraphics[width=0.5 \textwidth]{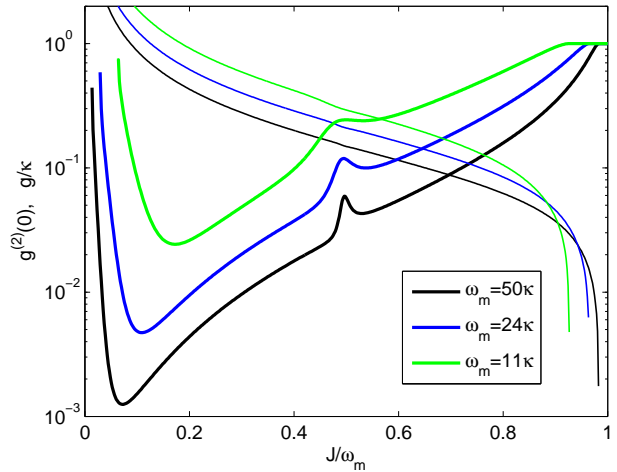}
\caption{\label{fig3} Thick lines: minimal value of $g^{(2)}(0)$ obtained by varying $g$, plotted as a function of $J$. Thin lines: corresponding value of $g$ for which the minimum occurs.}
\end{figure}

\begin{figure}
\includegraphics[width=0.5 \textwidth]{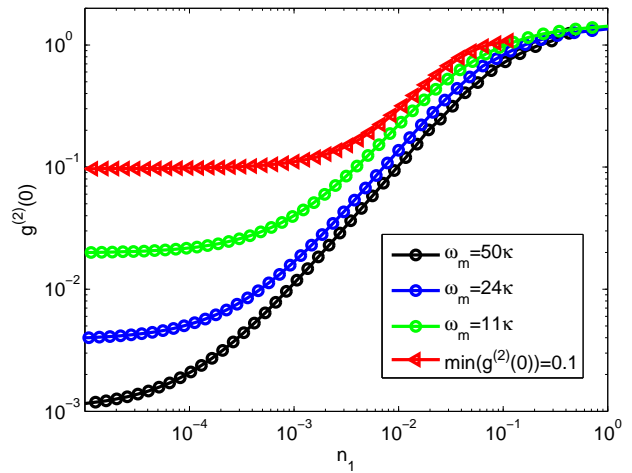}
\caption{\label{fig4} $g^{(2)}(0)$ as a function of the average photon occupation of the first mode $n_1$. Circles: parameters $g$ and $J$ set to the absolute minimum of $g^{(2)}(0)$ in Fig. \ref{fig3}. Triangles: parameters $g$ and $J$ corresponding to $\mbox{min}(g^{(2)}(0))=0.1$ in Fig. \ref{fig3} (points corresponding to different values of $\omega_m$ sit on the same curve). The rightmost points required up to $N_{ph}=22$ for convergence.}
\end{figure}

In order to quantify the effectiveness of the UPB mechanism, we have minimized the value of $g^{(2)}(0)$ by varying the optomechanical coupling $g$. Fig. \ref{fig3} displays $\mbox{min}(g^{(2)}(0))$ (thick lines), and the corresponding value of $g$ (thin lines), for varying $J$ and for the three values of $\omega_m$ considered. The absolute minimum of $g^{(2)}(0)$ is reached respectively for $J=1.9\kappa$ ($\omega_m=11\kappa$), $J=2.6\kappa$ ($\omega_m=24\kappa$), and $J=3.6\kappa$ ($\omega_m=50\kappa$), always significantly smaller than $\omega_m$, with the corresponding optimal value of $g\simeq1.1\kappa$ for the three cases. A small feature at $2J=\omega_m-\Delta_2$ arises because of resonant enhancement of the nonlinearity in coupled OMSs \cite{Ludwig2012,Stannigel2012}, but is scarcely significant in the present weak optomechanical coupling case. We see that, for the three mechanical frequencies, $g^{(2)}(0)\sim0.1$ can be achieved for $g\sim0.1\kappa$, corresponding to $\Delta_g<10^{-3}\kappa$, and a somewhat larger value of the optical coupling $J$. The increase in $\mbox{min}(g^{(2)}(0))$ with decreasing $\omega_m/\kappa$ is a consequence of the interferential nature of UPB. In fact, as suggested by the scheme proposed in Fig. \ref{fig2}, the well resolved sideband limit corresponds to having a small relative uncertainty in the phase of an excitation pathway involving phonon creation and destruction processes, which is in turn necessary for the occurrence of destructive interference.

To assess the expected rate of emission of antibunched photons, we study how $g^{(2)}(0)$ varies as the driving field amplitude $\epsilon$ is increased. Fig. \ref{fig4} displays $g^{(2)}(0)$ as a function of the average photon number in the first mode $n_1=\langle\hat{a}^\dagger_1\hat{a}_1\rangle$. Circles are computed for values of $g$ corresponding to the three absolute minima of $g^{(2)}(0)$ in Fig. \ref{fig3}, while squares correspond to values of $g$ and $J$ selected to have $g^{(2)}(0)=0.1$ in Fig. \ref{fig3}. This latter plot is identical for the three values of $\omega_m$, showing that $g^{(2)}(0)$ ultimately depends on the photon occupation only. More generally, we observe that $g^{(2)}(0)$ preserves its minimal value up to $n_1\sim10^{-2}$, contrarily to the Kerr case \cite{Liew2010}, where $g^{(2)}(0)$ decreases as a power law of $n_1$. This is due to the lower bound set by the phonon-assisted processes discussed above. For $n_1>10^{-2}$ the antibunching is progressively suppressed, and the coherent result $g^{(2)}(0)=1$ is recovered as the classical limit $n_1\to1$ is approached.

\begin{figure}[ht]
\includegraphics[width=0.5 \textwidth]{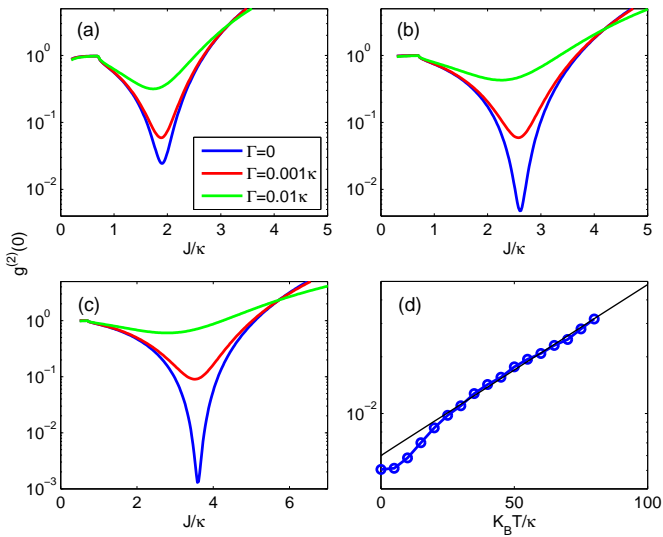}
\caption{\label{fig5} $g^{(2)}(0)$ vs $J$ for different values of the pure dephasing rate $\Gamma$, computed for (a) $\omega_m=11\kappa$, (b) $\omega_m=24\kappa$, and (c) $\omega_m=50\kappa$. For the three cases, $g$ was set to its optimal value corresponding to absolute minima of $g^{(2)}(0)$ in Fig. \ref{fig3}. (d) $g^{(2)}(0)$ vs temperature, as computed for $\omega_m=24\kappa$ and the optimal values of $g$ and $J$. Black line: thermally activated behaviour $g^{(2)}(0)\propto\exp(K_BT/2\omega_m)$. The rightmost points in (d) required up to $N_m=36$ for convergence.}
\end{figure}

Given the interferential origin of UPB, it is important to study how it is affected by pure dephasing induced by interactions with the environment. Pure dephasing typically arises from scattering processes of one photon with emission or absorption of one bath quantum \cite{Carmichael1999}. For mode one, they are described by the Hamiltonian $H_D=\sum_kg_k\hat{a}_1^\dagger\hat{a}_1(\hat{b}_k+\hat{b}_k^\dagger)$, where $\hat{b}_k$ are Bose operators describing modes of the thermal bath with given frequency $\omega_k$. A similar Hamiltonian holds for mode two. In the present case, terms of this kind can be interpreted as the optomechanical interaction of each photon mode with mechanical modes other than the one of interest. To lowest order, and within the Markov approximation, they are described as additional Lindblad terms to Eq. (\ref{eq:master})
\begin{equation}
\left.\frac{d \hat{\rho}}{dt}\right|_D=\frac{\Gamma}{2}\sum_{j=1,2}\mathcal{D}[\hat{a}_j^\dagger\hat{a}_j]\hat{\rho}\,,\label{eq:deph}
\end{equation}
where the magnitude of the rate $\Gamma$ -- assumed equal for the two modes -- is set by the polaron energy $g_k^2/\omega_k$. Fig. \ref{fig5} (a)-(c) show the variation of $g^{(2)}(0)$ as a function of $J$ for varying $\Gamma$, for $\omega_m=11\kappa$, $24\kappa$, and $50\kappa$ respectively. As expected for a phenomenon based on quantum interference, the UPB is suppressed by pure dephasing. The suppression is complete at a rate $\Gamma\sim0.01\kappa$, namely as soon as $\Gamma\gg \Delta_g$. This is easily understood, as $\Delta_g$ quantifies the phase shift required for exact destructive interference between excitation pathways to the two-photon state. In typical OMSs, the mechanical mode under study is tailored to maximize the optomechanical coupling $g$, among all other modes in the same frequency range. We thus expect that the condition $\Gamma\ll \Delta_g$ is fulfilled in most OMSs currently investigated.

Finally, Fig. \ref{fig5}(d) shows the variation of $g^{(2)}(0)$ with temperature, computed for the optimal UPB parameters in the case $\omega_m=24\kappa$.  The effect of temperature becomes relevant as soon as $K_BT\geq\omega_m$, where we see the onset of a thermally activated behaviour $g^{(2)}(0)\propto\exp(K_BT/2\omega_m)$. Accordingly, we expect $g^{(2)}(0)>0.1$ for $K_BT>150\kappa$, which is less than one decade below the value $K_BT\sim1000\kappa$ characterizing the mechanical mode in Ref. \onlinecite{Chan2012}.

The conditions required to observe UPB are radically different from those needed for the convetional mechanism. Firstly, the monochromatic driving field is not resonant with any of the eigenstates of the coupled system (light is absorbed by virtue of the finite linewidths). But most importantly, UPB requires a much smaller optomechanical nonlinearity $\Delta_g=g^2/\omega_m$, ranging from $10^{-4}\kappa$ to $10^{-1}\kappa$. In single OMSs \cite{Rabl2011}, CPB is observed for $\Delta_g/\kappa\gtrsim1$. In coupled OMSs on the other hand, the resonance condition $\omega_m\sim2J$ has been shown to enhance nonlinear effects also for $\Delta_g/\kappa\lesssim1$ \cite{Ludwig2012,Stannigel2012}. These resonant schemes however still rely on a strongly anharmonic spectrum and are thus bound to the additional condition of strong optomechanical coupling $g\gg\kappa$. Here, we have shown that UPB is achieved for the much less stringent condition $g/\kappa\sim0.1\div1$. These considerations suggest that UPB should be much more easily achievable in state-of-the art OMSs than CPB. Candidate systems must be in the well resolved sideband limit $\omega_m\gg\kappa$, thus excluding Bose-Einstein condensates of trapped ultracold atoms \cite{Brennecke2008,Gupta2007} for which $\omega_m$ is in the kHz range. To this purpose, the best strategy is probably to minimize the cavity loss rate $\kappa$ rather than to maximize $\omega_m$ at the expense of a lower optical quality factor \cite{Ding2011}. Ultimately, the systems closest to the UPB requirements are silica toroidal microresonators \cite{Verhagen2012} ($\omega_m/\kappa=11$, $g/\kappa=5\times10^{-4}$), optomechanical crystal nanobeams \cite{Chan2012,Davanco2012} ($\omega_m/\kappa=24$, $g/\kappa=5\times10^{-3}$), and micromechanical oscillators coupled to microwave cavities \cite{Teufel2011} ($\omega_m/\kappa=50$, $g/\kappa=10^{-3}$). The first two systems easily allow for a wide range of values of $J/\kappa$ \cite{Davanco2012,Grudinin2010}. Optomechanical crystal nanobeams \cite{Chan2012} in particular would require an improvement of the ratio $g/\kappa$ by a factor of 20, and cooling down to $T\sim1~\mbox{K}$. Another promising system is the Fabry-P\'erot GaAs/AlAs microcavity with distributed Bragg reflectors \cite{Fainstein2013} ($\omega_m/\kappa=1.7$, $g/\kappa=4\times10^{-3}$, $\kappa\sim2\pi\times12~\mbox{GHz}$), where UPB requirement could be easily reached for the current state-of-the-art loss rate $\kappa$ achievable in these systems \cite{Reitzenstein2007}. 

Quantum interference in coherently driven optical systems is a promising mechanism in view of the generation of quantum correlated states in presence of weak nonlinearities \cite{Liew2012,Liew2013,Carusotto2013}. Applied to OMSs, this paradigm might be the shortest route towards the controlled generation of quantum correlations between optical and mechanical  degrees of freedom.


\end{document}


\title{Unconventional photon blockade in coupled optomechanical systems:\\ Supplemental material}

\author{V. Savona}
\affiliation{Institute of Theoretical Physics, Ecole Polytechnique F\'{e}d\'{e}rale de Lausanne (EPFL), CH-1015 Lausanne, Switzerland}

\date{\today}

\maketitle

\begin{appendix}

\section{The numerical method}

The numerical calculations in the cases of large driving field amplitude $\epsilon$, or large temperature, required values of $N_{ph}$ up to 22 and of $N_m$ up to 36 respectively. Solving the stationary von Neumann equation for the density matrix $\hat{\rho}$ on such a large basis set (up to 800 states), as was done in the present work, requires an optimized numerical approach, that we describe below.

The starting problem is a homogeneous linear system of equations for the elements of the density matrix,
\begin{equation}
\mathcal{L}\hat{\rho}=0\,,\label{eq:Lrho}
\end{equation}
subject to the condition $\mbox{Tr}(\hat{\rho})=1$. Numerically, we need to ``vectorize'' the matrix $\hat{\rho}$, namely to express it as a column vector $\mbox{vec}(\hat{\rho})$ made by stacking the columns of $\hat{\rho}$ on top of one another. If $\rho$ is a $N\times N$ matrix (we omit the {\em hat} notation for all matrices below), then
\begin{equation}
\mbox{vec}(\rho)=[\rho_{1,1},\ldots\rho_{N,1},\rho_{1,2},\ldots\rho_{N,2},\ldots,\rho_{1,N},\ldots\rho_{N,N}]^T\,.
\label{eq:vecrho}
\end{equation}
To express the Lindblad superoperator in matrix form, to be used in this representation, we rely strongly on the property
\begin{equation}
\mbox{vec}(A \rho B)=\left(B^T\otimes A\right)\mbox{vec}(\rho)\,,\label{eq:vec}
\end{equation}
where $A$, $\rho$, and $B$ are $N\times N$ matrices and $\otimes$ denotes the Kronecker product. The matrix $B^T\otimes A$ is a $N^2\times N^2$ matrix. Indeed, all terms of the von Neumann equation are expressed as products of operators with the density matrix. In particular, the above identity gives
\begin{eqnarray}
\mbox{vec}([H,\rho])&=&\left(I\otimes H-H^T\otimes I\right)\mbox{vec}(\rho)\,,\label{eq:comm}\\
\mbox{vec}(2a_j\rho a^\dagger_j-a^\dagger_ja_j\rho-\rho a^\dagger_ja_j)
&=&\left(2(a^\dagger_j)^T\otimes a_j-I\otimes(a^\dagger_ja_j)-(a^\dagger_ja_j)\otimes I\right)\mbox{vec}(\rho)\,,\label{eq:lind}
\end{eqnarray}
where $I$ is the $N\times N$ identity matrix. From these identities, the whole Lindblad superoperator $\mathcal{L}$ can be cast in matrix form. 

To build the matrix corresponding to $\mathcal{L}$ acting on a truncated Fock-state space, the following procedure is applied. We start from the destruction operators on the truncated Fock-state space of each single mode (the creation operators are then simply obtained as the adjoint operators). We have
\begin{equation}
a_{N_{ph}}=\left(
\begin{array}{ccccc}
0&1&0&\cdots&0\\
0&0&\sqrt{2}&\cdots&0\\
\vdots&&\ddots&\ddots&\vdots\\
0&&\cdots&0&\sqrt{N_{ph}}\\
0&&\cdots&&0
\end{array}\right),\,\,\,\,\,
b_{N_{m}}=\left(
\begin{array}{ccccc}
0&1&0&\cdots&0\\
0&0&\sqrt{2}&\cdots&0\\
\vdots&&\ddots&\ddots&\vdots\\
0&&\cdots&0&\sqrt{N_{m}}\\
0&&\cdots&&0
\end{array}\right)\,,
\end{equation}
From these matrices we can build the corresponding matrices acting on the truncated Fock-state space of the multi-mode system. For example, for the case considered here of two photon and one phonon mode, the matrices of the corresponding destruction operators are built as
\begin{eqnarray}
a_1&=&a_{N_{ph}}\otimes I_{N_{ph}}\otimes I_{N_{m}}\,,\label{eq:a1}\\
a_2&=&I_{N_{ph}}\otimes a_{N_{ph}}\otimes I_{N_{m}}\,,\label{eq:a2}\\
b_2&=&I_{N_{ph}}\otimes I_{N_{ph}}\otimes b_{N_{m}}\,,\label{eq:b2}
\end{eqnarray}
where as before, $I_N$ is the $N\times N$ identity matrix. These matrices are now operating on a $(N_{ph}+1)^2(N_{m}+1)$-dimensional vector space. We can further reduce the computational task by restricting to the space generated by states $|n_1,n_2,n_m\rangle$ such that $n_1+n_2\le N_{ph}$ and $n_m\le N_m$. This step is justified by noticing that $n_1+n_2$ determines the total energy of the photon field, so that the truncation corresponds to retaining all states within a maximum total energy, responding to an intuitive perturbation criterion. As an example, for $N_{ph}=22$ and $N_m=2$ we have a Fock-state basis of $(N_{ph}+1)^2(N_{m}+1)=1587$ states, while only 828 states are left after the additional truncation procedure. Numerically, this truncation is obtained by building the vector $n_{ph}$ obtained from the diagonal of the matrix $a_1^\dagger a_1+a_2^\dagger a_2$, searching all elements of this vector for which $n_{ph}(j)>N_{ph}$ and removing from the matrices (\ref{eq:a1})-(\ref{eq:b2}) the corresponding $j$-th rows and columns.

From the matrices (\ref{eq:a1})-(\ref{eq:b2}), the Hamiltonian can be computed by simple matrix multiplications, following which, the matrix of the Lindblad superoperator $\mathcal{L}$ can be computed using the prescriptions (\ref{eq:comm})-(\ref{eq:lind}). Finally, the linear set of equations (\ref{eq:Lrho}), with the additional condition on the trace of the density matrix, can be solved by replacing one of the equations in the set by the condition $\mbox{Tr}({\rho})=1$ expressed in vectorized form. This is possible as we are assuming the existence of a solution to the stationary problem, and thus $\det(\mathcal{L})=0$. The resulting linear system is now inhomogeneous and can be solved by standard matrix inversion or by ad-hoc methods. We empirically found that the numerical accuracy of the result can depend on which line is replaced by the condition on the trace.

The Matlab\textsuperscript{\textregistered} computing language has powerful built-in linear algebra and matrix manipulation tools, that allow translating all the above steps into a very compact code. In particular, all matrices described above are highly sparse. This allows using the sparse-matrix representation available in Matlab\textsuperscript{\textregistered}, which results in a very moderate run-time memory requirement. The Kronecker product is also a built-in function in Matlab\textsuperscript{\textregistered}, and takes advantage of the sparse matrix representation. To solve the linear system, the built-in linear solver {\tt mldivide} is the fastest choice for small problems. It has the disadvantage however that the matrix is internally turned into its full-memory representation, thus rapidly hitting a memory limitation as the size of the problem grows. For larger problems, using one of the iterative methods available for sparse matrices is probably the best solution. We empirically found that the least-squares method, implemented by the Matlab\textsuperscript{\textregistered} function {\tt lsqr}, gives the best results for the problem treated in this work. In this way, the numerical solution of the stationary von Neumann problem for a given set of parameters required from less than one second -- for the smallest problems -- up to a few hours at most for the largest problems considered, and all computations could be carried out on a modern desktop PC with 32 GB of RAM.

\section{Weak pump limit}

In the limit of weak driving field, $\epsilon\to 0$, we can assume that only states with the lowest occupation numbers are significant for the occurrence of unconvetional photon blockade (UPB). In particular, we need to include at least states with one phonon and two photons, in order to account for the effective Kerr nonlinearity induced by the optomechanical coupling. We can then proceed in analogy with the analysis carried out by Bamba {\em et al.} \cite{Bamba2011} in the case of the simple Kerr nonlinearity. We thus express the steady state of the system as
\begin{equation}
|\psi\rangle=\sum_{n_1+n_2=0}^2\sum_{n_m=0}^1C_{n_1n_2n_m}|n_1n_2n_m\rangle\,.\label{eq:state}
\end{equation}
To determine the twelve coefficients $C_{n_1n_2n_m}$, in the steady state, in the limit of vanishing temperature and neglecting pure dephasing, we can write the Schr\"odinger equation for $|\psi\rangle$
\begin{equation}
i\frac{d|\psi\rangle}{dt}=\hat{H}|\psi\rangle\,,\label{eq:schr}
\end{equation}
and set $d|\psi\rangle/dt=0$. Here $\hat{H}$ is an effective non-hermitian Hamiltonian that accounts for the dissipation terms, defined as the original Hamiltonian
\begin{eqnarray}
\hat{H}&=&\Delta_1\hat{a}^\dagger_1\hat{a}_1+\Delta_2\hat{a}^\dagger_2\hat{a}_2+\omega_m\hat{b}^\dagger_2\hat{b}_2\notag\\
&-&J(\hat{a}^\dagger_1\hat{a}_2+\hat{a}^\dagger_2\hat{a}_1)
+g\hat{a}^\dagger_2\hat{a}_2(\hat{b}^\dagger_2+\hat{b}_2)+\epsilon(\hat{a}^\dagger_1+\hat{a}_1)\,,\label{eq:OmHam}
\end{eqnarray}
with the replacement $\Delta_j\to\Delta_j-i\kappa/2$ and $\omega_m\to\omega_m-i\gamma/2$. We further assume $\Delta_1=\Delta_2\equiv\Delta$ for simplicity (see discussion in the main article). 
Then, by replacing (\ref{eq:state}) into (\ref{eq:schr}), assuming steady state, and requiring that the the coefficient of each number state $|n_1n_2n_m\rangle$ vanish independently, we finally obtain the following set  of linear equations
\begin{eqnarray}
0&=&\Delta C_{100}+\epsilon C_{000} + \sqrt{2}\epsilon C_{200} + J C_{010}\,,\label{eq:1}\\
0&=&\Delta C_{010}+\epsilon C_{110} + J C_{100} + g C_{011}\,,\label{eq:2}\\
0&=&(\Delta+\omega_m)C_{011}+\epsilon C_{111}+J C_{101} + g C_{010}\,,\label{eq:3}\\
0&=&(\Delta+\omega_m)C_{101}+\epsilon C_{001}+\sqrt{2}\epsilon C_{201} +J C_{011}\,,\label{eq:4}\\
0&=&2\Delta C_{200}+\sqrt{2}\epsilon C_{100} +\sqrt{2}J C_{110}\,,\\
0&=&2\Delta C_{110}+\epsilon C_{010} +\sqrt{2}J (C_{200}+C_{020})+g C_{111}\,,\\
0&=&2\Delta C_{020}+\sqrt{2}J C_{110}+2g C_{021}\,,\\
0&=&(2\Delta+\omega_m)C_{021}+\sqrt{2}J C_{111} +2g C_{020}\,,\\
0&=&(2\Delta+\omega_m)C_{111}+\sqrt{2}J (C_{201}+C_{021}) +g C_{110}+\epsilon C_{011}\,,\\
0&=&(2\Delta+\omega_m)C_{201}+\sqrt{2}\epsilon C_{101} +\sqrt{2}J C_{111}\,,\label{eq:10}
\end{eqnarray}
Two additional equations, namely $\epsilon C_{100}=0$ and $\epsilon C_{101}+\omega_m C_{001}=0$, are irrelevant to the problem. The first one vanishes identically in the limit $\epsilon\to0$, while the second one determines the coefficient $C_{001}$ which however vanishes as $O(\epsilon^2)$. Furthermore, we can assume $C_{000}=1$, thus leaving the state unnormalized, which doesn't affect the value of $g^{(2)}(0)$. Finally notice that the leading order in the $\epsilon$-dependence of the various coefficients $C_{n_1n_2n_m}$ is $O(\epsilon^{n_1+n_2})$. Then, some terms can be neglected in Eqs. (\ref{eq:1})-(\ref{eq:10}), as they are of higher order in $\epsilon$. In particular, the terms $\sqrt{2}\epsilon C_{200}$ in (\ref{eq:1}), $\epsilon C_{110}$ in (\ref{eq:2}), $\epsilon C_{111}$ in (\ref{eq:3}), as well as the terms $\epsilon C_{001}+\sqrt{2}\epsilon C_{201}$ in (\ref{eq:4}) can be neglected. In this way, the matrix corresponding to the linear system is no longer hermitic as the original one, but the set of equations is now closed, and the coefficients $C_{n_1n_2n_m}$ are given by the leading order in $\epsilon$ only. A full analytical solution of the system obtained in this way is possible but rather cumbersome, and doesn't provide as much insight into the optimal values of parameters required for UPB, as the analogous system of equations studied in the Kerr case.\cite{Bamba2011} Such an insight can instead be gathered from approximate considerations, as follows.

The zero-delay two-photon correlation $g^{(2)}(0)$ in the first mode can be expressed as
\begin{eqnarray}
g^{(2)}(0)&=&\frac{\langle \hat{a}_1^\dagger\hat{a}_1^\dagger\hat{a}_1\hat{a}_1\rangle}{\langle \hat{a}_1^\dagger\hat{a}_1\rangle^2}\nonumber\\
&=&\frac{2|C_{200}|^2+2|C_{201}|^2}{(|C_{100}|^2+|C_{101}|^2)^2}\,.\label{eq:g2}
\end{eqnarray}
As all $C_{n_1n_2n_m}$ are homogeneous in $\epsilon^{n_1+n_2}$, this expression does not depend on $\epsilon$, as expected in the limit $\epsilon\to 0$.
Expression (\ref{eq:g2}) already clarifies the occurrence of a lower bound in $g^{(2)}(0)$, as discussed in the main article. It is determined by the presence of the one-phonon term in the numerator -- which is instead absent in the corresponding equation for the simple Kerr model.\cite{Bamba2011} There is no optimal choice of the parameters that makes the two terms in the numerator of (\ref{eq:g2}) vanish simultaneously. In particular, the coefficient $C_{201}$ never vanishes. We can prove this statement to lowest order in $g/\omega_m$. Let us divide Eqs. (\ref{eq:1})-(\ref{eq:10}) by $\omega_m$ and solve them iteratively in orders of $g/\omega_m$. Then, coefficients $C_{n_1n_2n_m}$ with $n_m=0$ only depend on even powers of $g/\omega_m$, while those with $n_m=1$ depend on odd powers of the same parameter. More precisely,
\begin{eqnarray}
C_{200}&=&\frac{\Delta^2}{\sqrt{2}(\Delta^2-J^2)^2}+\frac{g^2J^2(4\Delta^2\Delta_1+J^2\Delta_2)}{2\Delta(\Delta^2-J^2)^3(\Delta_1^2-J^2)}+O(g^4)\,,\label{eq:gsecond}\\
C_{201}&=&-\frac{g\sqrt{2}\Delta J^2}{(\Delta^2-J^2)^2(\Delta_1^2-J^2)}+O(g^3)\,,\label{eq:gfirst}
\end{eqnarray}
where all quantities are now normalized by $\omega_m$, we have assumed $\gamma=0$ for simplicity, and we have defined $\Delta_1\equiv\Delta+1$ and $\Delta_2\equiv2\Delta+1$. It is now clear that the first-order term in (\ref{eq:gfirst}) never vanishes unless either $\kappa$ or $J$ vanish. This is enough to prove analytically the existence of a lower bound on $g^{(2)}(0)$ in the optomechanical UPB mechanism occurring for $g/\omega_m\ll1$. Casting Eq. (\ref{eq:gsecond}) in a form with a common denominator instead, we can derive the conditions for vanishing $C_{200}$ by equating to zero the numerator of such expression. This gives
\begin{equation}
2\Delta^3(\Delta^2-J^2)(\Delta_1^2-J^2)+g^2J^2(4\Delta^2\Delta_1+J^2\Delta_2)=0\,.
\end{equation}
We can set back $\Delta\to\Delta-i\kappa/2$, $\Delta_1\to\Delta+\omega_m$, $\Delta_2\to2\Delta+\omega_m$, and require the real and imaginary parts of this expression to vanish independently. Then, in the limit $\Delta,\,\kappa,\,g\ll J\ll\omega_m$, we obtain after straightforward algebra
\begin{eqnarray}
\frac{\Delta_{\text{opt}}}{\kappa}&=&\pm\frac{1}{2\sqrt{3}}\,,\\
\frac{g^2_{\text{opt}}}{\kappa\omega_m}&=&\pm\frac{2}{3\sqrt{3}}\frac{\kappa^2}{J^2}\,,
\end{eqnarray}
that coincide with the optimal conditions found in Ref. \onlinecite{Bamba2011}, if we assume an effective Kerr nonlinearity $U_{\text{eff}}=g^2/\omega_m$, as occurs in an optomechanical system.

\begin{figure}
\includegraphics[width=0.5 \textwidth]{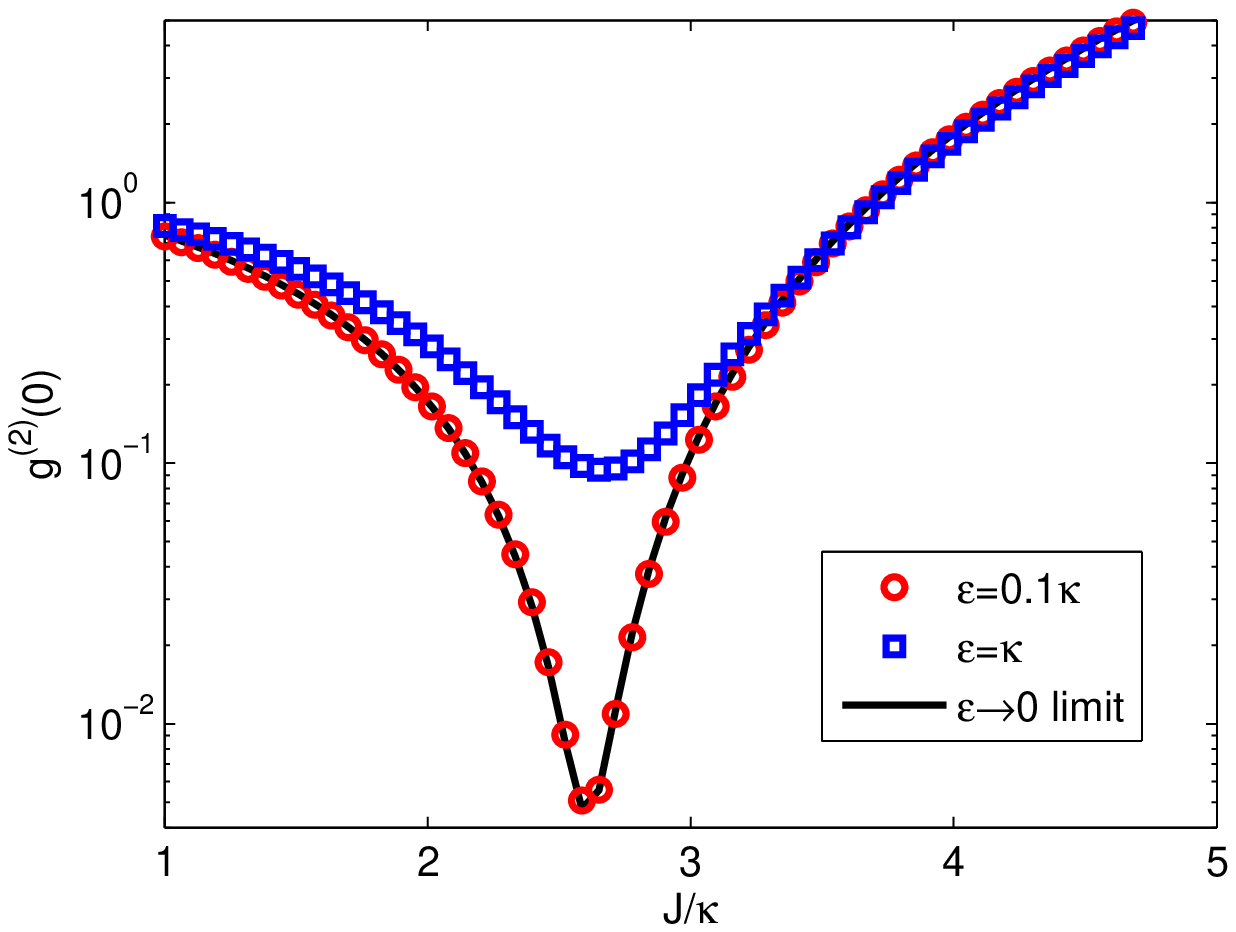}
\caption{\label{fig6} $g^{(2)}(0)$ as a function of $J$ computed for $\omega_m=24\kappa$, $g=1.16\kappa$, $\gamma=0.01\kappa$, $K_BT=0.1\kappa$. Line: analytical limit (\ref{eq:g2}). Circles: exact solution, $\epsilon=0.1\kappa$. Squares: exact solution, $\epsilon=\kappa$.}
\end{figure}

We conclude by solving the system (\ref{eq:1})-(\ref{eq:10}) numerically (keeping only the leading-order terms in $\epsilon$ in each equation as discussed above) and computing $g^{(2)}(0)$ from Eq. (\ref{eq:g2}), which holds in the limit $\epsilon\to0$. Fig. \ref{fig6} shows the quantity $g^{(2)}(0)$ computed from Eq. (\ref{eq:g2}) (line), from the fully converged solution of the stationary von Neumann problem for $\epsilon=0.1\kappa$ (circles), and $\epsilon=\kappa$ (squares). Clearly, the small-$\epsilon$ result is very well reproduced by the approximate model presented here, which is instead unable to account for the exact result in the case of stronger driving field $\epsilon=\kappa$.

\end{appendix}